\documentclass[superscriptaddress,secnumarabic,
amssymb,amsmath,nobibnotes,aps,prd,showkeys,showpacs,nofootinbib]{revtex4}%
\usepackage{graphicx}
\usepackage{epsf}
\usepackage{bm}
\usepackage{amsmath}
\usepackage{amsfonts}
\usepackage{amssymb}
\usepackage{epstopdf}
\usepackage{natbib}
\usepackage{color}%
\providecommand{\U}[1]{\protect\rule{.1in}{.1in}}
\newcommand{\be}{\begin{equation}}
\newcommand{\ee}{\end{equation}}

\newcommand{\mincir}{\raise
-3.truept\hbox{\rlap{\hbox{$\sim$}}\raise4.truept\hbox{$<$}\ }}
\newcommand{\magcir}{\raise
-3.truept\hbox{\rlap{\hbox{$\sim$}}\raise4.truept\hbox{$>$}\ }}

\begin{document}

\title{Inflation and late-time acceleration from a double-well potential with cosmological constant}



\author{Jaume de Haro\footnote{E-mail: jaime.haro@upc.edu}}
\affiliation{Departament de Matem\`atiques, Universitat Polit\`ecnica de Catalunya, Diagonal 647, 08028 Barcelona, Spain}
\author{Emilio Elizalde \footnote{E-mail:  elizalde@ieec.uab.es}}
\affiliation{ Consejo Superior de Investigaciones Cient\'{\i}ficas, ICE-CSIC and IEEC,
UAB Campus, C/ Can Magrans s/n, 08193 Bellaterra, Barcelona, Spain }


\thispagestyle{empty}

\begin{abstract}
 A model of a universe without big bang singularity is presented, which displays an early inflationary period ending just before
 a phase transition to a  kination epoch. The model produces enough heavy particles so as to reheat the universe at temperatures in the MeV regime. 
 After the reheating, it smoothly matches the standard $\Lambda$CDM scenario.
\end{abstract}

\vspace{0.5cm}


\maketitle

\section{ Introduction}

Starting from the celebrated surveys of type Ia supernovae \cite{1} as standard candles
and from the anisotropy findings in the power spectrum of the Cosmic Microwave Background (CMB) \cite{2}, which showed the evidence that the universe is undergoing a phase of 
accelerated expansion --that started quite recently in redshift scale-- the unified description of early time
inflation \cite{Inflation} and the current cosmic acceleration is one of the most attractive topics in cosmology nowadays.
Several possibilities, such as modified gravity \cite{odintsov}, quintessential inflation \cite{Spokoiny, pv}, the Horava-Lifshitz theory \cite{elizalde},
or entropic cosmology \cite{cai} have been put forward to actually perform this unification.

In \cite{hap}, following the spirit of quintessential inflation and
in order to unify it with the current cosmic acceleration, the authors have proposed a model where the  potential of the scalar field is a combination of a
double-well inflationary potential \cite{DWI} and a cosmological constant. This model provides  a (non geodesically past-complete)  background that could be obtained 
explicitly and allows to perform analytic calculations. It depicts a non-singular universe at finite  cosmic time (the big bang singularity being absent), which at early 
times exhibits an inflationary period followed by an abrupt phase transition to a stiff matter dominated (kination or deflationary) 
regime  \cite{Spokoiny, joyce},
able to produce, via 
gravitational pre-heating, a number of particles large enough in order to reheat the universe and to eventually match the standard $\Lambda$CDM model to high accuracy.

The aim of the present work is to study in detail some important aspects of this model and to explicitly calculate most relevant quantities, as the value of the reheating 
temperature and the precise time when it does occur. In fact, we will show that the obtained gravitational production of heavy, conformally coupled massive particles, with masses of the
order $m\sim 10^{12}$ GeV, leads to a
reheating temperature in the MeV regime. This
result is consistent with the  reheating temperature bounds coming from nucleosynthesis, which have been found to be of the order of $1$ MeV \cite{gkr}. Moreover, this low 
temperature prevents
a late time entropy production due to the decay of non-relativistic gravitational relics such as gravitinos or moduli particles \cite{kks}.

Furthermore, those heavy particles reach thermal equilibrium quite fast, namely in some $10^{-24}$ seconds after their production, leading to a relativistic plasma that 
is able to reheat the universe some $10^{-5}$ seconds after the phase transition.
We will also compare our model, in the aspects mentioned, with the pioneering one proposed by Peebles and Vilenkin \cite{pv}, with the result that, owing 
to the smoothness of the phase transition in that very popular model, reheating via heavy massive particle production leads, for particles with mass $m\sim 10^{12}$ GeV, to 
an abnormally small temperature of about $10^2$ eV. 

The present paper is structured as follows. In Sect.~2 we present the model and the corresponding background derived from it. Sect.~3 is devoted to the study of cosmological 
perturbations and it is there shown that our model fits well recent observational data. The reheating process is studied in detail in Sect.~3, where we prove, in particular, 
that our model leads to reheating temperatures in the MeV regime. Finally, in the last section  all the evolution of the inflation field from the phase transition to the present 
epoch is discussed.

The units used throughout the paper are $\hbar=c=1$.

\section{The model}

The gravitational model here considered is endowed with a cosmological constant, $\Lambda\equiv 4\lambda^4 M_{pl}^4$  ($\lambda$ being a dimensionless parameter different 
from zero and $M_{pl}$ the reduced Planck's mass), and a potential with the form (see \cite{hap} for a detailed description)
 \begin{eqnarray}\label{simple}
V(\varphi)=\left\{\begin{array}{ccc}
9\bar{\lambda}^4\left(\varphi^2-\frac{2}{3}M_{pl}^2\right)^2& \mbox{for} & \varphi<\varphi_E\\
4\lambda^4 M_{pl}^4& \mbox{for} & \varphi\geq \varphi_E,
\end{array}\right.
\end{eqnarray}
where $\varphi_E=-M_{pl}\sqrt{\frac{2}{3}}\sqrt{ \left(\frac{\lambda}{\bar\lambda}\right)^2+1}$, and $\bar{\lambda}\gg \lambda$ is another dimensionless parameter.
As we want the cosmological constant to dominate at present time, we have to impose
$\lambda^{4}M_{pl}^4\sim H_0^2 M_{pl}^2\Longrightarrow \lambda\sim 6\times 10^{-31}$
where we take into account that the current value of the Hubble parameter is $H_0\sim 6\times 10^{-61} M_{pl}$.

 An important property of the potential (\ref{simple}) is that the conservation equation
 \begin{eqnarray}
  \ddot{\varphi}+3H\dot{\varphi}+V_{\varphi}=0,
 \end{eqnarray}
has the following solution
 \begin{eqnarray}
  \varphi(t)=\left\{\begin{array}{ccc}
       -M_{pl}\sqrt{\frac{H(t)}{\sqrt{3}\bar{\lambda}^2M_{pl}}-\frac{2}{3}}&\mbox{for}& t<0\\
       &&\\
       -\frac{2}{3}M_{pl}\ln\left(\frac{\sqrt{3H^2(t)-2{\lambda^2M_{pl}^2}}+\sqrt{3}H(t)}{\sqrt{3H_E^2-2{\lambda^2M_{pl}^2}}+\sqrt{3}H_E}\right)+\varphi_E &\mbox{for}  & t>0,
                    \end{array}
\right.
 \end{eqnarray}
where $H_E=\frac{2M_{pl}}{\sqrt{3}}\left(2\bar{\lambda}^2+{\lambda^2} \right)$ and $H(t)$ is the background coming from the potential
\begin{eqnarray}\label{bbb}
H(t)=\left\{\begin{array}{ccc}
\frac{2M_{pl}}{\sqrt{3}}\left(\bar{\lambda}^2+ (\bar{\lambda}^2+{\lambda^2 })e^{-{8}{\sqrt{3}}\bar{\lambda}^2 M_{pl}t}\right)& \mbox{for} & t<0\\
& & \\
\frac{2\lambda^2 M_{pl}}{\sqrt{3}}\left(  \frac{\bar{\lambda}^2+\lambda^2 +\bar{\lambda}^2e^{-4\sqrt{3}\lambda^2 M_{pl} t}}{\bar{\lambda}^2+{\lambda^2}-\bar{\lambda}^2
e^{-4\sqrt{3}\lambda^2M_{pl}t}}         \right)            & \mbox{for} & t>0.
\end{array}\right.
\end{eqnarray}
For this background, the effective equation of state (EoS) parameter $w_{eff}(t)=-1-\frac{2\dot{H}}{3H^2}$  satisfies
\begin{eqnarray}
w_{eff}=\left\{\begin{array}{ccc}
-1 & \mbox{for} & t\ll -\frac{1}{\bar{\lambda}^2M_{pl}}\\
1 &\mbox{for} & 0<t\ll\frac{1}{\lambda^2M_{pl}}\\
-1 &\mbox{for} & t\gg\frac{1}{\lambda^2 M_{pl}}.
\end{array}\right.
\end{eqnarray}
This means that both at very early and at late times the universe is nearly de Sitter, thus providing a good description of the inflationary era, and of the current cosmic 
acceleration, respectively.
Moreover, after inflation the universe experiences a phase transition to a  kination phase, where heavy particles are produced in a sufficient amount in order to be able 
to reheat the universe, in accordance with the observational bounds.

Note also that the Hubble parameter, and thus the energy density, only diverge when $t\rightarrow -\infty$. This means that the big bang singularity
(understood as a divergence of the energy density at finite, early cosmic time) is not present in this model. Actually, in analogy with the so-called {\it little  rip} 
singularity where the EoS parameter tends asymptotically to $-1$ at future time (see, for instance, \cite{fls,beno}), we may argue
that, in our model, the universe starts in a {\it little bang}. Moreover,
 following the arguments given in \cite{bgv}, we will see that our background
is not past-complete. This can be easily realized, because for $t<0$ the scale factor in our model is given by
\begin{eqnarray}
 a(t)=a_E e^{{-\frac{\bar{\lambda}^2+\lambda^2}{12\bar{\lambda}^2}}e^{-{8}{\sqrt{3}}\bar{\lambda}^2 M_{pl}t}}e^{\frac{2M_{pl}}{\sqrt{3}}\bar{\lambda}^2t},
\end{eqnarray}
and thus,  the  maximum affine parameter
$
 \tilde{\lambda}_{max}\equiv \frac{1}{a_E}\int_{-\infty}^0 a(t)dt
$ is finite, meaning that  any backward-going null geodesic  has a finite affine length, i.e., it is past-incomplete.
The same happens with  massive particles moving along time-like geodesics, in this case let $p_0\not= 0$ be the three-momentum  at time $t=0$, then,  the maximum proper time
$\tau_{max}\equiv \int_{-\infty}^0 \frac{ma(t)}{\sqrt{m^2a^2(t)+p_0^2a^2(0)}}dt$ will also be finite.

In the same way, by choosing periodic potentials one can find a universe starting and ending in a de Sitter phase (see Eqs.~(26) and (68) of \cite{phpj}). The background actually leads to
a universe (although not geodesically past-complete)  where the energy density  never diverges.

Finally, note that  this analysis is at the classical level, while for energy densities at Planck scales the classical picture losses it sense. In fact, from the best of our knowledge
the only way to have a nonsingular universe which is geodesically complete is in the context of bouncing cosmologies \cite{bounces}, where in order to obtain a bounce one needs
to introduce
nonstandard matter fields \cite{nonstandard} or to go beyond General Relativity \cite{LQC}.

\section{Cosmological perturbations}
We first review some results obtained in \cite{hap} and start by
 introducing the  main  slow roll parameters \cite{btw},
\begin{eqnarray}\label{slowroll}
 \epsilon=-\frac{\dot{H}}{H^2}, \quad \eta=2\epsilon-\frac{\dot{\epsilon}}{2H\epsilon},
\end{eqnarray}
needed to calculate the  parameters associated with the power spectrum, namely the spectral index ($n_s$), its running ($\alpha_s$), and the
 ratio of tensor to scalalar perturbations  ($r$), defined by 
\begin{eqnarray}\label{8}
 n_s-1=-6\epsilon+2\eta, \quad \alpha_s\equiv \frac{d n_s}{d\ln(aH)}=\frac{H\dot{n}_s}{H^2 +\dot{H}},\quad
 r=16\epsilon.
\end{eqnarray}
Now,  taking de derivative with respect to the cosmic time of our background (\ref{bbb}), we see that, for $t<0$, we have
\begin{eqnarray}\label{9}
\dot{H}=-8\bar{\lambda}^2M_{pl}(\sqrt{3}H-2\bar{\lambda}^2M_{pl}),
\end{eqnarray}
 and, introducing a new variable $x\equiv \frac{8\sqrt{3}\bar{\lambda}^2M_{pl}}{H}$, the slow roll parameters (\ref{slowroll}) can be expressed, in an alternative way, as
\begin{eqnarray}\label{10}
 \epsilon=x\left(1-\frac{x}{12}\right),\quad \eta=\epsilon+\frac{x}{2}.
\end{eqnarray}
As a consequence, we find $
n_s-1=-3x+\frac{x^2}{3}$ and $r=16x-\frac{4x^2}{3}$, which depict a curve on the plane $(n_s,r)$ (Fig. $1$).

 Note that, since $n_s=0.9583\pm 0.0081$ \cite{Ade}, one has $x=\frac{9}{2}\left(1-\sqrt{1-\frac{4(1-n_s)}{27}}\right)\sim 10^{-2}$. Then, due to this small value of $x$
and using (\ref{10}), one has $\epsilon\cong x$. Moreover, we will have  $n_s-1\cong -3x$ and $r\cong 16x$, meaning that the curve in the plane $(n_s,r)$
is approximately the  straight line $n_s-1=-\frac{3}{16}r$, which is the same that one obtains for a quartic potential $V(\varphi)=\lambda_0\varphi^4$.
Effectively, for that potential we have to use the formulas 
\begin{eqnarray}\label{slowroll1}
\epsilon\cong \frac{M_{pl}^2}{2}\left(\frac{V_{\varphi}}{V} \right)^2=\frac{8M_{pl}^2}{\varphi^2}\quad
\mbox{and} \quad \eta\cong \frac{M_{pl}^2V_{\varphi\varphi}}{V}=\frac{12M_{pl}^2}{\varphi^2}.
\end{eqnarray}
A simple calculation leads finally to
\begin{eqnarray}
n_s-1=-\frac{24M_{pl}^2}{\varphi^2} \quad r=\frac{128M_{pl}^2}{\varphi^2} \Longrightarrow n_s-1=-\frac{3}{16}r. 
\end{eqnarray}

Another relevant quantity is given by   $N\equiv \int^{t_{end}}_t H(s)ds$, namely the number of e-folds from observable scales exiting the Hubble radius towards the end of inflation,  where
we choose, as usual, that inflation ends when $\epsilon=1$;
that is, when the Hubble parameter has the value $H_{end}=\frac{8\sqrt{3}\bar{\lambda}^2M_{pl}}{6(1-\sqrt{2/3})}$.
Then, since
$w_{eff}(H)=-1+\frac{2}{3}\epsilon$, the inflationary period will end when $w_{eff}(H_{end})=-\frac{1}{3}$, which implies that the universe will subsequently enter into a 
decelerating phase.
To calculate  the number of e-folds, we will perform the change of variable $H=H(t)$, thus obtaining the formula
$N(H)=-\int_{H_{end}}^H\frac{H}{\dot{H}}dH$,  which after inserting (\ref{bbb}) and (\ref{9}) in it leads to the following equation
\begin{eqnarray}
N(x)=\frac{1}{x}-\frac{1}{x_{end}}+\frac{1}{12}\ln\left(\frac{12-x}{12-x_{end}}\frac{x_{end}}{x} \right),
\end{eqnarray}
where $x_{end}=6(1-\sqrt{2/3})\cong 1.1010$ denotes the value of the parameter $x$ when inflation ends.

In order to check for the viability of the model, we consider
 the 2-dimensional marginalized confidence level on the plane $(n_s,r)$ in the presence of running, since our model includes it  (from the second formula of
 (\ref{8}) one easily deduces that $\alpha_s\cong -3x^2$), see Fig. $1$ --where  the black path corresponds 
 to the curve $((n_s(x),r(x))$ as coming from our model.
Planck2013 and Planck2015 observations respectively constraint  our parameter $x$ as follows.
 Planck2013 data \cite{Ade} at $95\%$ CL constraint 
the parameter $x$ to the interval $[0.0075, 0.0156]$ which means that the number of e-folds is bounded between $64\leq N(x)\leq 133$.
On the other hand, Planck2015 TT$+$low P data \cite{Planck} at $95\%$ CL constraint the parameter $x$ to be in the range $[0.0061, 0.0124]$, and hence, $80\leq N(x)\leq 163$.

 As we will show at the end of this section (see formula (\ref{efolds})),  nucleosynthesis bounds constrain the number of e-folds from $70$ to $80$, as a consequence
our model  matches perfectly with the Planck2013 data.

\begin{figure}
\includegraphics[height=0.37\textwidth,angle=0]{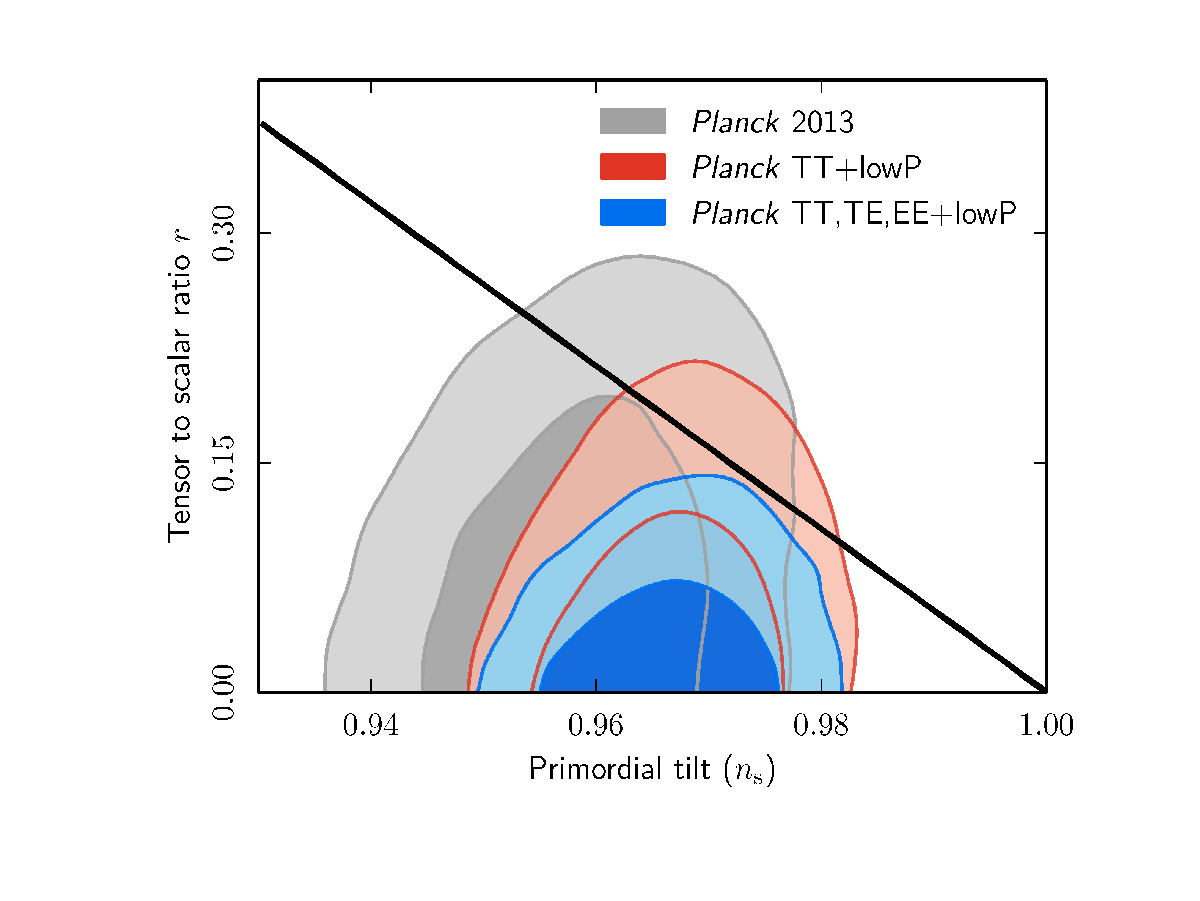}
\caption{Marginalized joint confidence contours for $(n_{\mathrm s} \,, r)$,
at the 68\,\% and 95\,\% CL, in the presence of running of the spectral indices. The black path corresponds to the curve $(n_s(x),r(x)))$ of our model
({Figure courtesy of the Planck2015 Collaboration}).
}
\label{fig:nsrr}
\end{figure}


To determine the value of the parameter $\bar\lambda$, one has to take into account the theoretical \cite{btw} and the observational \cite{bld} values of the power spectrum
when the pivot scale, namely $k_*$, crosses the Hubble radius
\begin{eqnarray}\label{power1}
 {\mathcal P}\cong \frac{H^2_*}{8\pi^2M_{pl}^2\epsilon}
 \cong \frac{24\bar{\lambda}^4}{\pi^2  x^3}
 \cong 2\times 10^{-9},
\end{eqnarray}
 where $H_*$ denotes the value of the Hubble parameter when the pivot scale leaves the Hubble radius.
Finally, using the values of $x$ in the range
{$[0.0075,0.0156]$,}
we  conclude that
\begin{eqnarray} \bar{\lambda}\sim  10^{-4}.
\end{eqnarray}

To get more in touch with observational data, we note that,  for a given $k$-mode,  we can write 
\begin{eqnarray}
\frac{k}{a_0H_0}
=e^{-N_k}\frac{H_k}{H_0}\frac{a_{end}}{a(0)}\frac{a(0)}{a_R}\frac{a_R}{a_M}\frac{a_M}{a_0}=
e^{-N_k}\frac{H_k}{H_0}\frac{a_{end}}{a(0)}\frac{\rho_R^{-1/12}\rho_M^{1/4}}{\rho^{1/6}(0)}\frac{a_M}{a_0},
\end{eqnarray}
where $R$ (resp. $M$) denotes the point when radiation (resp. matter) starts to dominate, 
 $N_k$ is the number of e-folds from   the $k$-mode exiting the Hubble radius to the end of inflation, $H_k$ is the value of the Hubble parameter when the $k$-mode leaves the Hubble radius,
and we have used the relation between the scale factor and the energy density in
the corresponding different phases
\begin{eqnarray}
\left(\frac{a(0)}{a_R}\right)^6=\frac{\rho_R}{\rho(0)}, \quad
\left(\frac{a_R}{a_M}\right)^4=\frac{\rho_M}{\rho_R}.
\end{eqnarray}
 Then, for modes in the current horizon scale $k=a_0H_0$, one has
\begin{eqnarray}
N_{hor}=\ln\left(\frac{H_{hor}}{H_0} \right)+\ln\left(\frac{a_{end}}{a(0)} \right)+
\frac{1}{4}\ln\left(\frac{\rho_M}{\rho_R} \right)+\frac{1}{6}\ln\left(\frac{\rho_R}{\rho(0)} \right)
+\ln\left(\frac{a_M}{a_0} \right).
\end{eqnarray}
If we assume, as usual, that inflation ends when the universe starts to decelerate ($\epsilon=1)$, this means
$\dot{H}=-H^2$; in our model, this will happen when $e^{-8\sqrt{3}M_{pl}\bar{\lambda}^2t_{end}}\cong 10$, thus, $H_{end}\cong \frac{20M_{pl}\bar{\lambda}^2}{\sqrt{3}}$ 
and $a_{end}=0.35 a(0)$, and, therefore,  $\ln\left(\frac{a_{end}}{a(0)} \right)\cong -1.04$.

Using  $\frac{a_0}{a_M}=3360$ \cite{rg}, one obtains
$\ln\left(\frac{a_M}{a_0} \right)\cong -8.12$. Further, from (\ref{power1})
we also have
\begin{eqnarray}
\ln\left(\frac{H_{hor}}{H_0}\right)\cong 130.75 +\frac{1}{2}\ln x.
\end{eqnarray}
Now, since the current temperature of the cosmic background is $T_0= 2.73$ K and the
conservation of entropy implies $T_M=\frac{a_0}{a_M}T_0\sim 9\times 10^3$ K $\sim 9\times 10^{-10}$ GeV,  using that
$\left(\frac{\rho_M}{\rho_R} \right)^{1/4}\cong \frac{T_M}{T_R}  $, we obtain that
\begin{eqnarray}
\frac{1}{4}\ln\left(\frac{\rho_M}{\rho_R} \right)\cong -20.83-\ln\left(\frac{T_R}{\mbox{GeV}}  \right).
\end{eqnarray}
Moreover, for our model it turns out that $\rho^{\frac{1}{4}}(0)\cong 2\bar{\lambda}M_{pl}\sim 2\times 10^{14}$ GeV. Then,
\begin{eqnarray}
\frac{1}{6}\ln\left(\frac{\rho_R}{\rho(0)} \right)=\frac{2}{3}\left(-32.57+\ln\left(\frac{T_R}{\mbox{GeV}}  \right)
\right).
\end{eqnarray}

Finally, collecting  all the results above, it follows that
we obtain
\begin{eqnarray}\label{efolds}
N_{hor}\cong 79.04+\frac{1}{2}\ln x-\frac{1}{3}\ln\left(\frac{T_R}{\mbox{GeV}}  \right),
\end{eqnarray}
what means that if the reheating temperature --with the purpose to ensure the success of nucleosynthesis--  needs to belong in the range between $10^9$ GeV and $1$ MeV, then
 the number of e-folds must lie between $70$ and $80$. In particular, when the reheating temperature is of the order of $10$ MeV --the scale we obtain
 if reheating is due to the creation of heavy particles with masses of about $10^{12}$ GeV during the phase transition--
 the number of e-folds of the universe expansion in our model, from  modes in the current horizon scale exiting the Hubble radius towards the end of inflation, is approximately
 $78$.

We observe that this number of e-folds is larger than the one obtained if it is assumed that there is not a substantial drop of energy density during the last stages of 
inflation ($\rho_{hor}\cong\rho_{end}$)
and that the universe reheats immediately after the end of it ($\rho_R\cong\rho_{end}$). Then, a simple calculation leads to
$N_{hor}\cong 68.25+ \frac{1}{4}\ln\left(\frac{\rho_{hor}}{M_{pl}^4}\right)\cong 64.6+ \frac{1}{4}\ln x\cong 62$ \cite{ll}.  In fact, from those results we can see that such
assumptions are  unjustified, because in our model
the drop of energy density plus a  reheating temperature compatible with nucleosynthesis leads to an increase of between  $8$ and $12$ e-folds.
Moreover, there is a key difference between standard inflation, where the potential has a minimum and particle production is due to the oscillations of the inflaton, and 
quintessential inflation, because in the first case it is usually assumed that from the end of inflation onto reheating the universe is matter dominated, leading to 
the formula \cite{ll}
\begin{eqnarray}
N_{hor}\cong 64.6+ \frac{1}{4}\ln \epsilon+\frac{1}{4}
\ln\left(\frac{\rho_{hor}}{\rho_{end}} \right)+ \frac{1}{12}
\ln\left(\frac{\rho_{R
}}{\rho_{end}} \right),
\end{eqnarray}
which gives, for admissible reheating temperatures, between $50$ and $60$ e-folds. However, in quintessential inflation after the phase transition to reheating the universe 
is stiff matter dominated, leading to the expression (recall that in our model $x\cong \epsilon$)
\begin{eqnarray}
N_{hor}\cong 64.6+ \frac{1}{4}\ln \epsilon+ \ln \left(\frac{a_{end}}{a(0)} \right)+\frac{1}{4}
\ln\left(\frac{\rho_{hor}}{\rho(0)} \right)-\frac{1}{12}
\ln\left(\frac{\rho_{R
}}{\rho(0)} \right),
\end{eqnarray}
which, we have showed, gives between $70$ and $80$ e-folds.
The main difference lies in the sign of the last term: in the first expression is positive, which means that the number of e-folds decreases, while in the second one it is 
negative, and thus,  the number of e-folds will grow.

A final remarks are in order:
Here we have calculated the number of e-folds for modes in the current horizon scale $k=a_0H_0$. However, if one chooses modes, as in the \cite{Ade}, in 
the scale of $k=0.005$ $\mbox{Mpc}^{-1}$, then
    the number of e-folds is reduced in $5$ units, and thus, for our model, it will lie between $65$ and $75$. This 
    result agrees with Planck2013 data (see figure 1), because as we have already explained,  at $95.5 \%$ C.L., for our model, Planck2013 data constrains 
    the number of e-folds  to lie between $64$ and $133$.
     

\section{The reheating process}

When dealing with the so-called  inflationary non-oscillatory models \cite{fkl}, i.e, models where the potential does not have an absolute minimum, preheating occurs due to a sudden
 phase transition from an inflationary phase to another one. There, the breakdown of adiabaticity leads to the production of particles coupled to gravity. For 
 instance, in \cite{ford}  the production of massless nearly conformally coupled particles originated in a sudden transition from inflation to the radiation era is studied.

Here, we will discuss the production of heavy massive $\chi$-particles ($m\gg \bar{\lambda}^2 M_{pl}$) conformally coupled to gravity coming from a phase transition to a 
 kination regime (see \cite{he} for details).

 Working in Fourier space, the dynamical equation of the $\chi$-particles is the same as an harmonic oscillatory
\begin{eqnarray}\chi_k''+\omega^2_k(\tau)\chi_k=0,
\end{eqnarray}
where the derivatives are taken with respect to the conformal time $\tau$ and the time dependent frequency of the $k$-mode is given by $\omega_k(\tau)=\sqrt{k^2+m^2a^2(\tau)}$.

In this case, since for our model (\ref{bbb}) the second derivative of the Hubble parameter is nearly continuous at the 
transition point (the difference between the second derivative  immediately before and after the phase transition being of
order $\lambda^2\bar{\lambda}^4 M_{pl}^3\sim 10^{-77} M_{pl}^3$), to take into account particle production during the adiabatic regimes,
one has to use the second-order WBK solution  (the first order WKB solution only contains first order derivatives of the 
Hubble parameter that, for our models, are always continuous, meaning that using this approximation it is impossible to take into account particle production) with the purpose 
to approximately define, before  the transition time,  the vacuum modes \cite{Haro}
\begin{eqnarray}\label{modes}
\chi_{2,k}^{WKB}(\tau)\equiv
\frac{1}{\sqrt{2W_{2,k}(\tau)}}e^{-{i}\int^{\tau}W_{2,k}(\bar{\tau})d\bar{\tau}},
\end{eqnarray}
 where the analytic expression of  $W_{2,k}$ was calculated in \cite{Bunch}
\begin{eqnarray}\label{wkb}
W_{2,k}=
\omega_k-\frac{m^2a^4}{4\omega_k^3}(\dot{H}+3H^2)+\frac{5m^4a^6}{8\omega_k^5}H^2+
\frac{m^2a^6}{16\omega_k^5}(\dddot{H}+15\ddot{H}H+10\dot{H}^2+86\dot{H}H^2+60H^4)\nonumber\\
-\frac{m^4a^8}{32\omega_k^7}(28\ddot{H}H+19\dot{H}^2+394\dot{H}H^2+507H^4)+\frac{221m^6a^{10}}{32\omega_k^9}(\dot{H}+3H^2)H^2-\frac{1105m^8a^{12}}{128\omega_k^{11}}H^4.
\end{eqnarray}

Note that, near the transition time, the adiabatic condition $\frac{\omega_{k}'}{\omega^2_{k}}\ll 1$ (the derivative is taken with respect to the conformal time)
is fulfilled, because one has
\begin{eqnarray}
 \frac{\omega_{k}'}{\omega^2_{k}}=\frac{m^2a^3H}{\omega^3_{k}}\sim \frac{H}{m}\sim \frac{\bar{\lambda}^2M_{pl}}{m}\ll 1,
\end{eqnarray}
what justifies the use of (\ref{modes}) to approximate the vacuum modes near the transition time.  However, at the transition time the positive and negative 
frequencies mix, and after the abrupt phase transition the vacuum modes become approximately
\begin{eqnarray}\label{modes1}
 \alpha_k\chi_{2,k}^{WKB}(\tau)+\beta_k(\chi_{2,k}^{WKB})^*(\tau),
\end{eqnarray}
where $\alpha_k$ and $\beta_k$ are Bogoliubov coefficients.

Then, impossing the continuity of the first derivative of (\ref{modes}) and (\ref{modes1}) at the transition time, one obtains the system 
\begin{eqnarray}
 \left\{\begin{array}{ccc}
         \chi_{2,k}^{WKB}(\tau_E^-)&=&\alpha_k\chi_{2,k}^{WKB}(\tau_E^+)+\beta_k(\chi_{2,k}^{WKB})^*(\tau_E^+)\\
         \frac{d}{d\tau}\chi_{2,k}^{WKB}(\tau_E^-)&=&\alpha_k\frac{d}{d\tau}\chi_{2,k}^{WKB}(\tau_E^+)+\beta_k\frac{d}{d\tau}(\chi_{2,k}^{WKB})^*(\tau_E^+),
        \end{array}
\right.
\end{eqnarray}
where $\chi_{2,k}^{WKB}(\tau_E^-)$ (resp. $\chi_{2,k}^{WKB}(\tau_E^+)$) is the value of $\chi_{2,k}^{WKB}(\tau)$ before (resp. after) the phase transtion time $\tau_E$.
Simple algebra shows that the $\beta_k$-Bogoliubov coefficient is given by (see \cite{Haro,he})
\begin{eqnarray}
\beta_k=\frac{{\mathcal W}[\chi_{2,k}^{WKB}(\tau_E^-),\chi_{2,k}^{WKB}(\tau_E^+)]}
{{\mathcal W}[(\chi_{2,k}^{WKB})^*(\tau_E^+),\chi_{2,k}^{WKB}(\tau_E^+)]},
\end{eqnarray}
where
${\mathcal W}[f(\tau_E^-), g(\tau_E^+)]\equiv f(\tau_E^-)g'(\tau_E^+)-f'(\tau_E^-)g(\tau_E^+)$ is the Wronskian of the functions $f$ and $g$ at time $\tau_E$.

\vspace{0.25cm}


Taking into account that the only discontinuous term in (\ref{wkb}) is $ \frac{m^2a^6\dddot{H}}{16\omega_k^5}$,
it is not difficult to show that  the squared modulus of the Bogoliubov coefficient  is given by
\begin{eqnarray}
 |\beta_k|^2\cong \frac{m^4a_E^{12}\left(\dddot{H}(0^+)-\dddot{H}(0^-)\right)^2}{1024(k^2+m^2a^2(0))^6},
\end{eqnarray}
where $\dddot{H}(0^-)$ (resp. $\dddot{H}(0^+)$), is the value of the second derivative of the Hubble parameter before (after) the phase transition. Recall that, for our model,
this quantity is discontinuous at the transition time.

From here, using that $\int_0^{\infty}\frac{y^2}{(y^2+1)^6}dy= \frac{7!!}{10!!}\frac{\pi}{2}$, the number density of produced particles and their energy density are, respectively,
 \cite{Birrell}
\begin{eqnarray}
 n_{\chi}(t)\equiv\frac{1}{2\pi^2 a^3(t)}\int_0^{\infty}k^2|\beta_k|^2 dk\sim
 5 \bar{\lambda}^{6}M_{pl}^3 \left(\frac{\bar{\lambda}^{2}M_{pl}}{m}\right)^5\left(\frac{a(0)}{a(t)} \right)^3,
 \rho_{\chi}(t)\equiv\frac{1}{2\pi^2 a^4(t)}\int_0^{\infty}\omega_k(t)k^2|\beta_k|^2 dk
 \sim mn_{\chi}(t).
\end{eqnarray}
Since these heavy particles are far from thermal equilibrium, they will decay into lighter particles, which will interact through multiple scattering, and thus,
redistribute their energies to achieve a relativistic
plasma phase in thermal equilibrium (for a more detailed explanation, see \cite{37,38}).


\vspace{0.5cm}

To calculate the moment when thermalization occurs, we will use the thermalization process depicted in \cite{38}, where the cross section is given by
$\sigma={\alpha^3}\rho_{\chi}^{-\frac{1}{2}}(0)$, with $\alpha^2\sim 10^{-3}$. Then, the thermalization rate is
\begin{eqnarray}
 \Gamma=\sigma n_{\chi}(0)=\alpha^3\left(\frac{n_{\chi}(0)}{m} \right)^{\frac{1}{2}}\sim \alpha^3\left(\frac{\bar{\lambda}^2M_{pl}}{m}\right)^3\bar{\lambda}^2M_{pl}.
\end{eqnarray}

Equilibrium is reached when $\Gamma\sim H(t_{eq})=\bar{\lambda}^2M_{pl}\left(\frac{a(0)}{a_{eq}}\right)^3$, obtaining
$\frac{a(0)}{a_{eq}}\sim \alpha  \frac{\bar{\lambda}^2 M_{pl}}{m}$. Thus, at the time of equilibrium,
the energy densities of the produced particles and background are, respectively,
\begin{eqnarray}
 \rho_{\chi}(t_{eq})\sim 5\alpha^3\bar{\lambda}^{8}\left(\frac{\bar{\lambda}^{2}M_{pl}}{m}\ \right)^{7}{M_{pl}^4}, \quad
 \rho(t_{eq})\sim 3\alpha^6 \bar{\lambda}^4 \left(\frac{\bar{\lambda}^2M_{pl}}{m}\ \right)^{6}M_{pl}^4.
\end{eqnarray}

After this thermalization, the relativistic plasma evolves as $\rho_{\chi}(t)=\rho_{\chi}(t_{eq})\left(\frac{a_{eq}}{a(t)} \right)^4$, and the background evolves as
$\rho(t)=\rho(t_{eq})\left(\frac{a_{eq}}{a(t)} \right)^6$, because we are in the kination regime. Reheating is obtained when both energy densities are of the same order, and
this will happen when $\frac{a_{eq}}{a_{R}}\sim \sqrt{\frac{\rho_{\chi}(t_{eq})}{\rho(t_{eq})}}$. Thus, we
obtain a reheating temperature of the order
\begin{eqnarray}
 T_R\sim \rho_{\chi}^{\frac{1}{4}}(t_{eq})\sqrt{\frac{\rho_{\chi}(t_{eq})}{\rho(t_{eq})}}\sim \alpha^{-\frac{3}{4}}\bar{\lambda}^4 \left(\frac{\bar{\lambda}^2M_{pl}}{m}\
 \right)^{\frac{9}{4}}M_{pl}
 \sim 10^3\left(\frac{\bar{\lambda}^2M_{pl}}{m}\ \right)^{\frac{9}{4}} \mbox{ GeV}.
\end{eqnarray}
As an example, if we consider heavy particles with mass
$m\sim 10^{12}$ GeV (recall that the condition $m\gg \bar{\lambda}^2 M_{pl}$, implies $m\gg 10^{10}$ GeV), the temperature reduces to  $T_R\sim 3\times 10$ MeV, that is,
we obtain a temperature in the MeV regime.

Note that there are different thermalization processes. For example, in \cite{37} the authors propose the following thermalization rate $\Gamma=\alpha^2 n_{\chi}^{\frac{1}{3}}$, which has been used recently in \cite{hap} in order to calculate the reheating temperature. In fact, for that process the reheating temperature is approximately 
 \begin{eqnarray}
 T_R\sim 
 \sim 10^4\left(\frac{\bar{\lambda}^2M_{pl}}{m}\ \right)^{\frac{19}{8}} \mbox{ GeV},
\end{eqnarray}
which for masses satisfying $m\sim 10^{12}$ GeV, is of the order of $10^2$ MeV.

\vspace{0.5cm}

Finally, since we have the explicit form of the Hubble parameter in (\ref{bbb}), we can calculate the time when reheating occurs, via the equality
\begin{eqnarray}
 3M_{pl}^2H^2(t_R)\sim \alpha^{-3} \bar{\lambda}^{16}\left(\frac{\bar{\lambda}^2M_{pl}}{m}\ \right)^{{9}}M_{pl}^4.
\end{eqnarray}
Using the approximation $e^{-4\sqrt{3}\lambda^2M_{pl}t_R}\cong 1-4\sqrt{3}\lambda^2M_{pl}t_R$,  one can see that $H(t_R)\sim t_R^{-1}$, and thus
\begin{eqnarray}
 t_R\sim \alpha^{\frac{3}{2}}\bar{\lambda}^{-8}\left(\frac{\bar{\lambda}^2M_{pl}}{m}\ \right)^{-\frac{9}{2}}\frac{1}{M_{pl}}.
\end{eqnarray}
In the case of particles with mass $m\sim 10^{12}$ GeV, using that $t_{pl}\sim 5\times 10^{-44} s$, we obtain that the time from the phase transition
to the end of reheating is of the order of
\begin{eqnarray}
 t_R\sim 5\times 10^{38}\frac{1}{M_{pl}}\sim 2\times 10^{-5}s.
\end{eqnarray}
In the same way we can calculate when the equilibrium will occur, via the identity $\rho(t_{eq})=3H^2(t_{eq})M_{pl}^2$, thus obtaining
\begin{eqnarray}
 H(t_{eq})\sim \alpha^3\bar{\lambda}^2\left(\frac{\bar{\lambda}^2M_{pl}}{m}\ \right)^{3}{M_{pl}}\Longrightarrow t_{eq}\sim
 \alpha^{-3}\bar{\lambda}^{-2}\left(\frac{\bar{\lambda}^2M_{pl}}{m}\ \right)^{-3}\frac{1}{M_{pl}},
\end{eqnarray}
which, in the case  $m\sim 10^{12}$ Gev, becomes $t_{eq}\sim 10^{-24} s$.

A last important remark is in order. In \cite{pv} the authors proposed the following model
\begin{eqnarray}\label{peebles}
 V(\varphi)=\left\{\begin{array}{ccc}
   \lambda_ 1(\varphi^4+M^4)&\mbox{for}& \varphi\leq 0\\
   \frac{\lambda_1 M^8}{\varphi^4+ M^4}&\mbox{for}& \varphi\geq  0,
                   \end{array}\right.
\end{eqnarray}
where, to match with observations, the dimensionless constant $\lambda_1$ has to be of order $10^{-14}$.

Since the derivatives of the potential are continuous up to order three included, it turns out that the first discontinuity in the
Hubble parameter appears in the fifth derivative. In fact, $\frac{d^5H(t_E^+)}{dt^5}-\frac{d^5H(t_E^-)}{dt^5}\sim
\frac{\lambda_1\dot{\varphi}^4(t_E)}{M_{pl}^2}$, where $t_E$ is the transition time.

In order to calculate this quantity, we need do some considerations. First of all, for our model \cite{1} the Hubble parameter at the transition time is $H_E\cong
\frac{4\bar{\lambda}^2M_{pl}}{\sqrt{3}}$,  and since 
its value when the pivot scale leaves the Hubble radius is 
$H_*\cong\frac{8\sqrt{3}\bar{\lambda}^2M_{pl}}{\epsilon}$,   the  energy drops by approximately $\frac{H_E}{H_*}\cong \frac{\epsilon}{6}$. Second, we will assume that 
for the Peebles-Vilenkin model one has an energy drop of the same order. Then, being so that when the pivot scale leaves the Hubble radius all the energy density is potential one, we will
have $H_*=\sqrt{\frac{V(\varphi)}{3}}\cong \frac{8\sqrt{\lambda_1}}{\sqrt{3}\epsilon}M_{pl} $, where we did approximate the potential by $V(\varphi)=\lambda_1\varphi^4$, and we have used the first equation of (\ref{slowroll1}). We thus obtain that at the transition time the Hubble parameter is approximately 
$H_E\sim \frac{4\sqrt{\lambda_1}}{3\sqrt{3}}M_{pl} $, for such model. Finally, since at the transition time all the energy density is kinetic, we will have
$\dot{\varphi}(t_E)=\sqrt{6}H_{E}M_{pl}\sim \sqrt{\lambda_1} M^2_{pl}$, what means that 
$\frac{d^5H(t_E^+)}{dt^5}-\frac{d^5H(t_E^-)}{dt^5}\sim \lambda_1^3 M_{pl}^6$.

On the other hand, using the WKB approximation at a higher order, it follows that the square modulus of the $\beta$-Bogoliubov coefficient is
 \begin{eqnarray}
 |\beta_k|^2\sim \frac{m^4a_E^{16}\lambda_1^{6}M_{pl}^{12}}{(k^2+m^2a^2_E)^8},
\end{eqnarray}
what means that in the Peebles-Vilenkin model the number density of particles produced at the transition time, namely $n_{\chi }^{PV}(t_E)$, is given by
\begin{eqnarray}
 n_{\chi }^{PV}(t_E)\sim \lambda_1^{\frac{3}{2}} \left(\frac{\sqrt{\lambda_1} M_{pl}}{m}\right)^9 M_{pl}^3.
\end{eqnarray}
Note the relation $\lambda_1=4\bar{\lambda}^4$; thus, to compare both models one has to choose
$\sqrt{\lambda_1}\sim 10^{-8}$. In this situation,
dealing with particles of mass $m\sim 10^{12}$ GeV, we find $n_{\chi }^{PV}(t_E)\sim 10^{-42} M_{pl}^3$ which is $8$ orders smallers than the number density of particles obtained in
our model, and thus, we conclude that the Peebles-Vilenkin model leads to an {\it abnormally} small reheating temperature. Actually, following the same steps as above one finds, for that
model, a reheating temperature of the order $T_R^{PV}\sim \alpha^{-\frac{3}{4}}\lambda_1\left(\frac{\sqrt{\lambda_1} M_{pl}}{m}\right)^{\frac{19}{4}}M_{pl}
\sim  10^{3} \left(\frac{\sqrt{\lambda_1} M_{pl}}{m}\right)^{\frac{19}{4}}$ GeV. This means that for particles with mass $m\sim 10^{12}$ GeV, the reheating temperature, in the
Peebles-Vilenkin model would be of the order $T_R^{PV}\sim 3\times 10^{-7}$ GeV $\sim 3\times 10^2$ eV $\sim 3\times 10^{6}$ K, which is in fact a very low temperature.


For further comparison, even lower reheating temperatures appear, in the above sense, in the models of \cite{Spokoiny}, where some very smooth potentials are chosen in order to describe
universes with an early inflationary period and
a late-time acceleration, leading indeed to very small reheating temperatures.
In all that papers, however, it is never assumed that reheating is due to particle production of heavy particles conformally coupled to gravity, on the contrary, it is arguably 
produced there by
light particles not conformally coupled with gravity (recall that massless conformally coupled particles are never at play), and previous results about reheating obtained in
\cite{ford,Damour, Giovannini} are used. The problem is that these results are model dependent and that, in all of them, an abrupt phase transition is crucially assumed --in order to
get the necessary high amount of particle production-- what invalidates their application to
models with a smooth phase transition (see the discussion in \cite{hap}).


\section{Evolution after reheating}
At the reheating time the kinetic energy of the field will be of the same order of the energy density of the universe $\dot{\varphi}^2(t_R)\sim \rho(t_R)$, because 
the potential energy
is given by $V(\varphi)=4\lambda^4M_{pl}^4\sim 10^{-121} M_{pl}^4$, which is smaller than the energy density at reheating.

To find the evolution of the scalar field after reheating, note that when the relativistic plasma starts to dominate,
the Hubble parameter is given by
\begin{eqnarray}
 H(t)=\frac{H_R}{1+2(t-t_R)H_R},
\end{eqnarray}
where $H_R\sim \frac{\sqrt{\rho(t_R)}}{M_{pl}}\sim  \frac{\dot{\varphi}(t_R)}{M_{pl}}$,
what means that
the scalar field satisfies the equation
\begin{eqnarray}
 \ddot{\varphi}+ \frac{3H_R}{1+2(t-t_R)H_R}\dot{\varphi}=0,
\end{eqnarray}
which solution is
\begin{eqnarray}
 \dot{\varphi}(t)=\frac{\dot{\varphi}(t_R)}{\left(1+2(t-t_R)H_R\right)^{\frac{3}{2}}}.
\end{eqnarray}
Since matter decays as $a^{-3}$ and radiation as $a^{-4}$,
the universe enters into a matter domination regime at $t_M$, which can be calculated as follows.
As we have seen in Sect.~3, due to the adiabatic regime after the phase transition, the temperature at the beginning of matter domination
will be $T_M\cong 9\times 10^3$ K. Then, using that $\rho(t_M)\sim H^2(t_M)M_{pl}^2\sim \frac{M_{pl}^2}{t_M^2}\sim T_M^4$, one gets
$t_M\sim \frac{M_{pl}}{T_M^2}\sim 10^{54} \frac{1}{M_{pl}}\sim 10^{10}$ s $\sim 3\times 10^{2}$ y.

As a consequence, for times after $t_M$,  the field satisfies the equation
\begin{eqnarray}\label{matter}
 \ddot{\varphi}+ \frac{6H_M}{2+3(t-t_M)H_M}\dot{\varphi}=0,
\end{eqnarray}
where $H_M\cong \frac{H_R}{1+2(t_M-t_R)H_R}$ is the value of the Hubble parameter at the beginning of the matter domination epoch.
The solution of (\ref{matter}) is
\begin{eqnarray}
 \dot{\varphi}(t)=\frac{\dot{\varphi}(t_M)}{\left(2+3(t-t_M)H_M\right)^{{2}}},
\end{eqnarray}
where now $\dot{\varphi}(t_M)\cong \frac{\dot{\varphi}(t_R)}{\left(1+2(t_M-t_R)H_R\right)^{\frac{3}{2}}}$.

At some given time $t_{\Lambda}$ the cosmological constant starts to dominate; this happens when
\begin{eqnarray}
 H^2(t_{\Lambda})M_{pl}^2\sim \lambda^4M_{pl}^4 \Longrightarrow (t_{\Lambda}-t_M)^{-1}\sim \lambda^2M_{pl},
\end{eqnarray}
that is, when $t_{\Lambda}\sim t_M+10^{20}s\sim t_{M}+3\times 10^{12} y.$
Further, when the cosmological constant dominates the fied equation will be
\begin{eqnarray}\label{lamda}
 \ddot{\varphi}+ \sqrt{12}\lambda^2M_{pl}\dot{\varphi}=0,
\end{eqnarray}
which solution is
\begin{eqnarray}
 \dot{\varphi}(t)=\dot{\varphi}(t_{\Lambda})e^{-\sqrt{12}\lambda^2M_{pl}(t-t_{\Lambda})},
\end{eqnarray}
where
\begin{eqnarray}
 \dot{\varphi}(t_{\Lambda})=\frac{\dot{\varphi}(t_{M})}{(2+(t_{\Lambda}-t_M)H_M)^2}\sim \frac{\dot{\varphi}(t_{M})\lambda^4M_{pl}^2}{H_M^2}=
 \frac{\dot{\varphi}(t_{R})\lambda^4M_{pl}^2}{H_R^2}(1+2(t_M-t_R)H_R)^{\frac{1}{2}}\nonumber\\
 \sim \frac{\lambda^4M_{pl}^4}{\rho_{\chi}^{\frac{1}{2}}(t_R)}(1+2(t_M-t_R)H_R)^{\frac{1}{2}}
 \ll \frac{\lambda^3M_{pl}^3}{\rho_{\chi}^{\frac{1}{4}}(t_R)},
\end{eqnarray}
where we have use that $t_M-t_R\ll t_{\Lambda}-t_M\sim \lambda^{-2} \frac{1}{M_{pl}}.$

From this result, we conclude that at the time $t_{\Lambda}$ the ratio of the kinetic to the potential energy is bounded
\begin{eqnarray}
 R\equiv\frac{\dot{\varphi}^2(t_{\Lambda})}{\lambda^4 M_{pl}^4}\ll \frac{\lambda^2 M_{pl}^2}{\rho_{\chi}^{\frac{1}{2}}(t_R)}\sim
 \alpha^{\frac{3}{2}}\left(\frac{\lambda}{\bar{\lambda}^4} \right)^2\left(\frac{\bar{\lambda}^2M_{pl}}{m}\ \right)^{-\frac{9}{2}}
 \sim 2\times 10^{-31}\left(\frac{\bar{\lambda}^2M_{pl}}{m}\ \right)^{-\frac{9}{2}}.
\end{eqnarray}
We realize that for heavy particles with mass $m\sim 10^{12}$ GeV, this ratio satisfies $R\ll 2\times 10^{-22}\ll 1$, meaning that at  present time the kinetic energy of
the field is sub-dominant and it will be the potential one which will drive the universe evolution.

\section{Conclusions}
\label{discuss}
We have considered a model that unifies  inflation with the current cosmic acceleration via a single scalar field whose potential is the combination of a double well inflationary
potential and a cosmological constant. The model provides a  background that is free from a big bang singularity, and which can be studied analytically, as we have here shown. It 
exhibits an early inflationary period where, for
observable modes, the universe inflates for a number of  $70$ to $80$ e-folds.
This number seems large as compared with the usual range of e-fold values used  to discard inflationary models. The reason behind is that, in standard inflation, where the potential as a minimum of energy and particle production is due to the oscillations of the inflaton, the universe evolves, from the end of
inflation to reheating,  as if it were  matter dominated.
However, in quintessential inflation, from the phase transition all the way to  reheating, the universe evolves instead as if it were driven by an stiff fluid, and it is this difference that is responsible for an increase in the number of e-folds in favor of non-oscillatory  models.

At the end of the inflationary period, the universe experiences a sudden phase transition
to a  kination phase, where heavy massive particles are created which reheat the universe some $10^{-5}$ seconds after they appear. At that point the universe enters into a radiation regime that finishes, as we have calculated, some $300$ years after the phase transition,  the universe then becoming matter dominated. This domination lasts, as has been here shown, for about $3\times 10^{12}$ years after the beginning of the matter domination stage, until the cosmological constant starts to rule the universe evolution, what is still happening today.

\vspace{0.5cm}

{\it Acknowledgments.}  We would like to thank professor Yifu Cai for his comments, in particular concerning  the past-incompleteness of our model. This investigation has been 
supported in part by MINECO (Spain), projects MTM2014-52402-C3-1-P and FIS2013-44881, by I-LINK1019 from CSIC, and by the CPAN Consolider Ingenio Project.

\end{document}